\documentclass[10pt,twocolumn,letterpaper]{article}

\usepackage{iccv}
\usepackage{times}
\usepackage{epsfig}
\usepackage{graphicx}
\usepackage{amsmath}
\usepackage{amssymb}
\usepackage{nicefrac}
\usepackage{mathtools}
\usepackage{adjustbox}
\usepackage{amsmath}

\usepackage{caption}
\usepackage{colortbl}
\usepackage{booktabs}
\usepackage{tikz}
\usepackage{pgfplots}
\usetikzlibrary{spy,calc}
\usepackage{array}
\newcolumntype{x}[1]{>{\centering\arraybackslash\hspace{0pt}}p{#1}}

\definecolor{sh_gray}{rgb}{0.84,0.84,0.84}
\definecolor{sh_gray2}{rgb}{1,0.89,0.75}
\definecolor{color3}{rgb}{0.95,0.95,0.95}
\definecolor{color4}{rgb}{0.96,0.96,0.86}
\definecolor{color5}{rgb}{0.90,0.90,0.90}

\newlength{\Oldarrayrulewidth}

\usepackage[pagebackref=true,breaklinks=true,letterpaper=true,colorlinks,bookmarks=false]{hyperref}

\newif\ifblackandwhitecycle
\gdef\patternnumber{0}

\pgfkeys{/tikz/.cd,
    zoombox paths/.style={
        draw=orange,
        very thick
    },
    black and white/.is choice,
    black and white/.default=static,
    black and white/static/.style={
        draw=white,
        zoombox paths/.append style={
            draw=white,
            postaction={
                draw=black,
                loosely dashed
            }
        }
    },
    black and white/static/.code={
        \gdef\patternnumber{1}
    },
    black and white/cycle/.code={
        \blackandwhitecycletrue
        \gdef\patternnumber{1}
    },
    black and white pattern/.is choice,
    black and white pattern/0/.style={},
    black and white pattern/1/.style={
            draw=white,
            postaction={
                draw=black,
                dash pattern=on 2pt off 2pt
            }
    },
    black and white pattern/2/.style={
            draw=white,
            postaction={
                draw=black,
                dash pattern=on 4pt off 4pt
            }
    },
    black and white pattern/3/.style={
            draw=white,
            postaction={
                draw=black,
                dash pattern=on 4pt off 4pt on 1pt off 4pt
            }
    },
    black and white pattern/4/.style={
            draw=white,
            postaction={
                draw=black,
                dash pattern=on 4pt off 2pt on 2 pt off 2pt on 2 pt off 2pt
            }
    },
    zoomboxarray inner gap/.initial=5pt,
    zoomboxarray columns/.initial=2,
    zoomboxarray rows/.initial=2,
    subfigurename/.initial={},
    figurename/.initial={zoombox},
    zoomboxarray/.style={
        execute at begin picture={
            \begin{scope}[
                spy using outlines={%
                    zoombox paths,
                    width=\imagewidth / \pgfkeysvalueof{/tikz/zoomboxarray columns} - (\pgfkeysvalueof{/tikz/zoomboxarray columns} - 1) / \pgfkeysvalueof{/tikz/zoomboxarray columns} * \pgfkeysvalueof{/tikz/zoomboxarray inner gap} -\pgflinewidth,
                    height=\imageheight / \pgfkeysvalueof{/tikz/zoomboxarray rows} - (\pgfkeysvalueof{/tikz/zoomboxarray rows} - 1) / \pgfkeysvalueof{/tikz/zoomboxarray rows} * \pgfkeysvalueof{/tikz/zoomboxarray inner gap}-\pgflinewidth,
                    magnification=3,
                    every spy on node/.style={
                        zoombox paths
                    },
                    every spy in node/.style={
                        zoombox paths
                    }
                }
            ]
        },
        execute at end picture={
            \end{scope}
     \gdef\patternnumber{0}
        },
        spymargin/.initial=0.5em,
        zoomboxes xshift/.initial=1,
        zoomboxes right/.code=\pgfkeys{/tikz/zoomboxes xshift=1},
        zoomboxes left/.code=\pgfkeys{/tikz/zoomboxes xshift=-1},
        zoomboxes yshift/.initial=0,
        zoomboxes above/.code={
            \pgfkeys{/tikz/zoomboxes yshift=1},
            \pgfkeys{/tikz/zoomboxes xshift=0}
        },
        zoomboxes below/.code={
            \pgfkeys{/tikz/zoomboxes yshift=-1},
            \pgfkeys{/tikz/zoomboxes xshift=0}
        },
        caption margin/.initial=0.4ex,
    },
    adjust caption spacing/.code={},
    image container/.style={
        inner sep=0pt,
        at=(image.north),
        anchor=north,
        adjust caption spacing
    },
    zoomboxes container/.style={
        inner sep=0pt,
        at=(image.north),
        anchor=north,
        name=zoomboxes container,
        xshift=\pgfkeysvalueof{/tikz/zoomboxes xshift}*(\imagewidth+\pgfkeysvalueof{/tikz/spymargin}),
        yshift=\pgfkeysvalueof{/tikz/zoomboxes yshift}*(\imageheight+\pgfkeysvalueof{/tikz/spymargin}+\pgfkeysvalueof{/tikz/caption margin}),
        adjust caption spacing
    },
    calculate dimensions/.code={
        \pgfpointdiff{\pgfpointanchor{image}{south west} }{\pgfpointanchor{image}{north east} }
        \pgfgetlastxy{\imagewidth}{\imageheight}
        \global\let\imagewidth=\imagewidth
        \global\let\imageheight=\imageheight
        \gdef\columncount{1}
        \gdef\rowcount{1}
        
    },
    image node/.style={
        inner sep=0pt,
        name=image,
        anchor=south west,
        append after command={
            [calculate dimensions]
            node [image container,subfigurename=\pgfkeysvalueof{/tikz/figurename}-image] {\phantomimage}
            node [zoomboxes container,subfigurename=\pgfkeysvalueof{/tikz/figurename}-zoom] {\phantomimage}
        }
    },
    color code/.style={
        zoombox paths/.append style={draw=#1}
    },
    connect zoomboxes/.style={
    spy connection path={\draw[draw=none,zoombox paths] (tikzspyonnode) -- (tikzspyinnode);}
    },
    help grid code/.code={
        \begin{scope}[
                x={(image.south east)},
                y={(image.north west)},
                font=\footnotesize,
                help lines,
                overlay
            ]
            \foreach \x in {0,1,...,9} {
                \draw(\x/10,0) -- (\x/10,1);
                \node [anchor=north] at (\x/10,0) {0.\x};
            }
            \foreach \y in {0,1,...,9} {
                \draw(0,\y/10) -- (1,\y/10);                        \node [anchor=east] at (0,\y/10) {0.\y};
            }
        \end{scope}
    },
    help grid/.style={
        append after command={
            [help grid code]
        }
    },
}

\newcommand\phantomimage{%
    \phantom{%
        \rule{\imagewidth}{\imageheight}%
    }%
}
\newcommand\zoombox[2][]{
    \begin{scope}[zoombox paths]
        \pgfmathsetmacro\xpos{
            (\columncount-1)*(\imagewidth / \pgfkeysvalueof{/tikz/zoomboxarray columns} + \pgfkeysvalueof{/tikz/zoomboxarray inner gap} / \pgfkeysvalueof{/tikz/zoomboxarray columns} ) + \pgflinewidth
        }
        \pgfmathsetmacro\ypos{
            (\rowcount-1)*( \imageheight / \pgfkeysvalueof{/tikz/zoomboxarray rows} + \pgfkeysvalueof{/tikz/zoomboxarray inner gap} / \pgfkeysvalueof{/tikz/zoomboxarray rows} ) + 0.5*\pgflinewidth
        }
        \edef\dospy{\noexpand\spy [
            #1,
            zoombox paths/.append style={
                black and white pattern=\patternnumber
            },
            every spy on node/.append style={#1},
            x=\imagewidth,
            y=\imageheight
        ] on (#2) in node [anchor=north west] at ($(zoomboxes container.north west)+(\xpos pt,-\ypos pt)$);}
        \dospy
        \pgfmathtruncatemacro\pgfmathresult{ifthenelse(\columncount==\pgfkeysvalueof{/tikz/zoomboxarray columns},\rowcount+1,\rowcount)}
        \global\let\rowcount=\pgfmathresult
        \pgfmathtruncatemacro\pgfmathresult{ifthenelse(\columncount==\pgfkeysvalueof{/tikz/zoomboxarray columns},1,\columncount+1)}
        \global\let\columncount=\pgfmathresult
        \ifblackandwhitecycle
            \pgfmathtruncatemacro{\newpatternnumber}{\patternnumber+1}
            \global\edef\patternnumber{\newpatternnumber}
        \fi
    \end{scope}
}

\usepackage[acronym,shortcuts]{glossaries}

\usepackage{setspace}

\usepackage[pagebackref=true,breaklinks=true,letterpaper=true,colorlinks,bookmarks=false]{hyperref}

\iccvfinalcopy 

\def\iccvPaperID{3956} 

\ificcvfinal\fi
\usepackage{cuted}
\usepackage{capt-of}

\makeglossaries

\newacronym{SISR}{SISR}{single image super-resolution}
\newacronym{PSNR}{PSNR}{peak signal-to-noise ratio}
\newacronym{MSE}{MSE}{mean squared error}
\newacronym{CNN}{CNN}{convolutional neural network}
\newacronym{ESPCN}{ESPCN}{efficient sub-pixel convolutional neural network}
\newacronym{DRCN}{DRCN}{deeply-recursive convolutional network}
\newacronym{ResNet}{ResNet}{residual network}
\newacronym{GAN}{GAN}{generative adversarial network}
\newacronym{LR}{LR}{low-resolution}
\newacronym{HR}{HR}{high-resolution}
\newacronym{SR}{SR}{super-resolution}
\newacronym{MOS}{MOS}{mean opinion score}
\newacronym{SSIM}{SSIM}{structural similarity}
\newacronym{PSNR-HVS}{PSNR-HVS}{peak signal-to-noise ratio - human visual system}
\newacronym{MS-SSIM}{MS-SSIM}{multi-granularity structural similarity}
\newacronym{SRGAN}{SRGAN}{super-resolution generative adversarial network}

\usepackage{subcaption}

\begin{document}

\title{Extreme Low-Light Imaging with Multi-granulation Cooperative Networks}

\author{Keqi Wang$^{1}$, Peng Gao$^{2}$, Steven Hoi$^{3}$,Steven Hoi$^{3}$, Qian Guo$^{1}$, Yuhua Qian$^{1}$\\
$^{1}$Institute of Big Data Science and Industry, Shanxi
University\\
$^{2}$The Chinese University of Hong Kong\\
$^{3}$Singapore Management University\\
{\tt\small 76190504@qq.com}\\
{\tt\small 1155102382@link.cuhk.edu.hk}\\
{\tt\small czguoqian@163.com}\\
{\tt\small jinchengqyh@126.com}
}


\maketitle

\maketitle

\begin{abstract}
Low-light imaging is challenging since images may appear to be dark and noised due to low signal-to-noise ratio, complex image content, and the variety in shooting scenes in extreme low-light condition. Many methods have been proposed to enhance the imaging quality under extreme low-light conditions, but it remains difficult to obtain satisfactory results, especially when they attempt to retain high dynamic range (HDR). In this paper, we propose a novel method of multi-granulation cooperative networks (MCN) with bidirectional information flow to enhance extreme low-light images, and design an illumination map estimation function (IMEF) to preserve high dynamic range (HDR). To facilitate this research, we also contribute to create a new benchmark dataset of real-world Dark High Dynamic Range (DHDR) images to evaluate the performance of high dynamic preservation in low light environment. Experimental results show that the proposed method outperforms the state-of-the-art approaches in terms of both visual effects and quantitative analysis.

\end{abstract}

\section{Introduction}
\label{sec:Introduction}

The imaging equipment has been well-developed lately, but low-light imaging is still a challenging work.
Under the low-light condition,
the imaging approach by traditional image signal processing (ISP) techniques always suffers darkness and noise due to the low signal-to-noise ratio (SNR) of captured signal. One strategy is to prolong exposure time for obtaining clearer image. However, owning to camera shake or object motion, longer exposure time may induce blur, and thus this strategy is not applicable to the video shooting. Another strategy is to flash for capturing a brighter image. The main drawback of this strategy is that the image looks unnatural.
\begin{figure}[ht]
    \centering
  	\begin{tabular}{cccc}
 		
     	\includegraphics[width=1.5in]{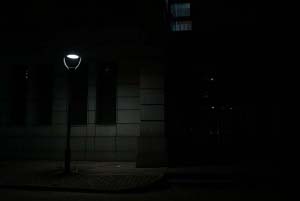}~~
        \includegraphics[width=1.5in]{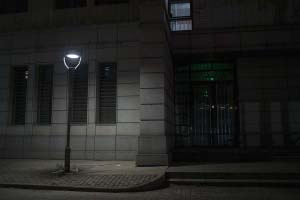}\\

        	\end{tabular}
        \begin{tabular}{cx{4cm}cx{6cm}}
	       ~~~~~(a)~Sony A7R3 & ~~~~~~~~~~~~~~~~~~(b)~Photoshop\\
        \end{tabular}

     	\begin{tabular}{cccc}
     	\includegraphics[width=1.5in]{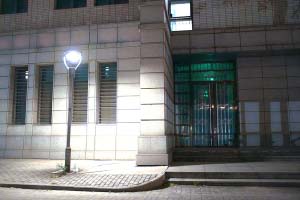} ~~
     	\includegraphics[width=1.5in]{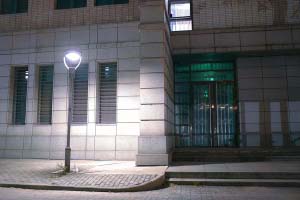}\\
     	\end{tabular}
             \begin{tabular}{cx{4cm}cx{6cm}}
	       ~~~~(c)~LTS & ~~~~~~~~~~~~~~~~~~~~~~~~~~~(d)~RMCN \\
        \end{tabular}

  \label{fig:IMEF}
\caption{Comparison of visual results by different approaches:
(a) Extreme low-light image shot by Sony A7R3.
(b) Photoshop Camera Raw automatic mode processing.
(c) LTS \cite{dark}.
(d) RMCN w/ RIMEF (ours).
Note that there are many noise in (b), over exposure and visible artifacts in (c).}
\label{fig:1}
\end{figure}
With the rapid development of deep learning, many computer vision tasks have been significantly advanced, such as image classification \cite{simonyan2015very}, face recognition \cite{cao2018pose-robust}, visual question answering \cite{gao2019dynamic, gao2018question, gao2019multi}, robust design \cite{lyu2019fastened}, and image style conversion \cite{Gatys2015A}. Imaging is also irreversible and in this process of converting raw data to JPEG, there is a huge information loss, which inspired researchers to substitute data-driven-based method for traditional image signal processing(ISP) algorithms.
Chen \etal \cite{dark} collected a large-scale dataset under extreme low-light conditions, which contains both indoor and outdoor images shot by two different cameras with different bayer color filter array (CFA), and a processing pipeline was proposed.

However, the above method has a large chromatic aberration and numerous corresponding details are missing in generated image, when shotting with severely limited illumination. To mitigate the above problems, a critical factor is the generative capacity of neural networks.
Many evidences indicate that performance can be improved with the increase of depth of the neural network, but the neural network structure should be designed cautiously to avoid network degradation.
He \etal \cite{He2015Deep} have reported this problem and addressed it by introducing residual learning.
This success may can be ascribed to improve the feature interaction between different layers of the neural network. Huang \etal \cite{Huang2017Densely} can also be seen as using this idea. Although the above methods have made some progress, in the view of granular computing, the above methods are defined by using single granulation instead of not optimal results.
And the idea of multi-granulation analysis has been successfully applied to feature extraction \cite{qian2010positive} and feature fusion \cite{qian2015fusing}.
To overcome this limitation, we inject the multi-granulation idea into the neural network algorithm, called multi-granulation cooperative network (MCN) with bidirectional information flow, which can be used to decompose a granularity into many small granularities.
Fusing all of the estimate on each small granularities together, the algorithm can get an optimal result.

In addition, how to preserve high dynamic range in low-light environment is another challenging question \cite{Chen_2019_ICCV, dark} (\figurename~\ref{fig:1}).
Although high dynamic range (HDR) problem has been studied for a long time by a large number of researchers, such as Ma \etal \cite{ma2015multi-exposure}.
However, their method is only applicable to JPEG images, not to RAW images in extreme low-light environment. To address the problem, a novel raw illumination map estimation function (RIMEF) is proposed, which preserves high dynamic range without multiple images and requires extremely small extra computation. Due to the lack of high dynamic range measurement data in extreme low-light environment, we collect a new dataset named ``Dark High Dynamic Range" (DHDR) to evaluate our method. To the best of our knowledge, this is the first method for preserving high dynamic range in low-light environment.
In short, the main contributions of this paper can be summarized as follows:

\begin{itemize}
    \item A new learning-based pipeline is proposed with multi-granulation cooperative network.
        Compared with other state-of-the-art methods, our method not only achieves higher PSNR and SSIM \cite{wang2004image}, but also has better visual effect.
    \item A novel raw illumination map estimation function is proposed, which can generate well-exposed sRGB images with the desired attributes (sharpness, color vividness, good contrast) in extreme low-light environment. Through a series of experiments, we verify the robustness of the proposed algorithm.
    \item A new Dark High Dynamic Range Dataset is collected, which is taken by Sony camera in extreme low-light high dynamic environment.

\end{itemize}
\section{Related Work}
In this section, we mainly discuss camera processing pipeline in Section~\ref{sec:Camera Processing Pipeline}, and related methods of low-light image enhancement in Section~\ref{sec:Low-light Image Enhancement}.

\subsection{Camera Processing Pipeline}
\label{sec:Camera Processing Pipeline}
For satisfying human perception, camera image signal processing (ISP)  \cite{ramanath2005color} algorithms could convert raw sensor data into image data accurately.
In general, digital camera is composed of preprocessing, white balance, demosaicking, color correction, gamma correction and post processing.
However, the ISP algorithms have limitations in special shooting scenarios such as shooting high-speed moving objects, shooting long-distance objects, shooting in low-light environment and so on.
Recently, the data-driven ISP algorithms have attracted increasing attention. Schwartz \etal \cite{schwartz2018deepisp} proposed a method to learn a mapping from low-light mosaiced image to well-lit images.
To apply machine learning to digital zoom, a new dataset and processing pipeline was proposed by \etal \cite{zhang2019zoom}. Chen \etal \cite{dark} introduced a pipeline for extreme low-light imaging from raw data to sRGB images.

\label{sec:pp}
\begin{figure*}[htb]
    \includegraphics[width=1\textwidth]{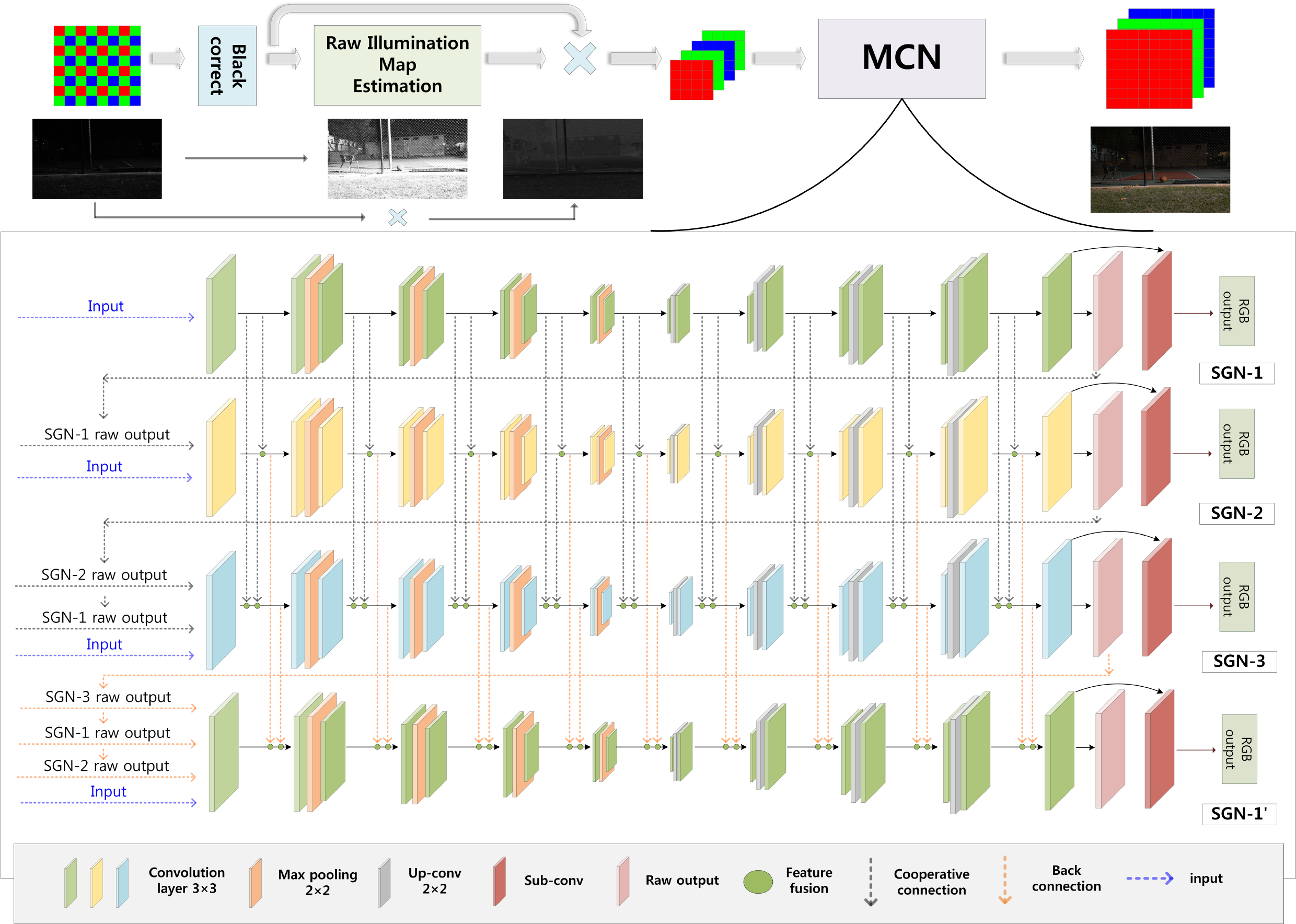}
    \caption{
    Processing pipeline.
    SGN-1 and SGN-1' are siamese networks.}
    \label{fig:coor_pipeline}
\end{figure*}

\subsection{Low-light Image Enhancement}
\label{sec:Low-light Image Enhancement}
Low-light imaging has been an open challenge in computer vision and image processing. Pictures captured under low-light conditions, especially under extreme low-light conditions, often have severe noises. Many well-known and classic enhancement methods are based on histogram equalization \cite{Lee2013Contrast,Pisano1998Contrast}.
However, these methods often cause over-saturation or under-saturation of certain areas of the enhanced image, and generate a lot of noise at the same time.
Some classic but more sophisticated methods are Retinex-based \cite{Park2017Low,Jobson2002A,Guo2017LIME}, which improve the image from two factors, one is reflection and the other is illumination, the enhanced image of this method usually looks unnatural and has an over-enhanced situation.
With the rapid development of deep learning, more and more researches have been done to enhance low-light image. The method of \cite{Lore2017LLNet} can learn to enhance and denoise adaptively through a variant of the stacked-sparse denoising autoencoder. The method of \cite{Shen2017MSR} combines convolutional neural networks with Retinex theory and makes great progress. Recently, Loh \etal \cite{Loh2018Getting} mainly focuses on object detection in low-light scenes and proposes the Exclusively Dark (ExDARK) dataset.

\section{Method}
In this section, we introduce the details of our proposed framework.
Firstly, we introduce our overall pipeline in Section~\ref{sec:processing_pipeline}.
Second, we give a detailed definition of multi-granulation cooperative network (MCN) and several specific examplese of it in Section~\ref{sec:munet}.
Then, we introduce a novel raw illumination map estimation function (RIMEF) in Section~\ref{sec:IMEF}.
Finally, we present our loss function in Section~\ref{sec:loss function}.
\subsection{Processing Pipeline}
\label{sec:processing_pipeline}
Under extreme low-light environment, the photons captured by the camera lens are extremely weak. After transforming raw image with 14-bit or higher into JPEG format generally with 8-bit, more details in the image will be lost. To keep more information, we choose raw data as the input of our pipeline.
First of all, in order to handle the existence of dark current, the black level correction is needed, and the pixel value will be scaled subsequently.
The difference between previous studies and ours is that the former generally scale pixel value by linear \cite{dark, Chen_2019_ICCV} patterns, while we do by designing the pattern using our proposed raw illumination map estimation function.
As a result, our method is able to generate better visual effect in high dynamic environment.
After the above operations, the Bayer arrays will be packed into $4$ channels and the X-Trans arrays will be packed into $9$ channels .
Finally we input the processed raw data into the neural network for learning its transfer to sRGB image. \figurename~\ref{fig:coor_pipeline} shows the overall pipeline of our method.
\subsection{Multi-granulation Cooperative Network}
\label{sec:munet}
In this subsection, we present the general framework of  multi-granulation cooperative networks (MCN), and introduce its two instantiations.


Inspired by the idea of multi-granulation and human cooperative intelligence, we propose a new network named multi-granulation cooperative network with bidirectional information flow (cooperative connection and back connection).
It contains multiple single-granularity networks (SGN) of various appearances. In this paper, our single-granularity network contains nine convolution blocks, each block, similar to the U-net, has two convolution layers. Note the output of single-granulation network has three channels, which can not be fed directly to the next network which requires a four- or nine-channel input. To utilize the output of the current SGN as the input of the rest networks, it is sufficient to add a raw output convolution layer to alter the channel number of the raw output to be same as that of the SGN input.
The last layer of each single-granulation network is a sub convolution layer.
The multi-granulation cooperative process pipeline is shown as follows:
\begin{equation}
h_{1,1} = \phi_{1,1}(h_{1,in}; w_{1,in},b_{1,in})
\end{equation}
where \emph{h$_{1,in}$} represents the input of the first single-granulation network,
\emph{$\phi$($\cdot$)} is the activate function and we adopt LReLU,
$w$ is the weight and $b$ is the bias,
\begin{equation}
h_{1,j+1} = \phi_{1,j}(h_{1,1},...,h_{1,j}; w_{1,j},b_{1,j}), j\in \{1,...,n\}
\end{equation}
where $j$ is the $j-th$ layer and $j\in \{1,...,n\}$ is the number of layers in single-granulation network,

\begin{equation}
h_{1,out} = \phi_{1,n+1}(h_{1,n+1},w_{1,n+1},b_{1,n+1})
\end{equation}
here $h_{1,out}$ is the output of the first SGN (SGN-1) before back connection,
an operation that sends each layer's feature of every but the first SGN to SGN-1, which can further improve the performance of multi-granulation cooperative network (\tablename~\ref{table:dsgr}).
It is noticeable that SGN-1 is a siamese network.

\begin{equation}
\begin{split}
h_{i,1} = \phi_{i,1}(f(&\alpha_{1,out} \times h_{1,out},...,\alpha_{i-1,out} \times h_{i-1,out}, \\
&h_{i,in}), w_{i,1},b_{i,1}),i\in\{2,...,n\}
\end{split}
\end{equation}
where $i$ is the $i-th$ SGN (SGN-$i$) and $i\in\{2,...,n\}$ stands for the order of a particular SGN,
$\alpha_{i,out}$ is a parameter to control the weight of $h_{i-1,out}$,
$f(\cdot)$ is the operation of feature fusion, which is used by cooperation connection or back connection. We will introduce some instantiations later in section~\ref{sec:Instantiations},

\begin{equation}
\begin{split}
h_{i,j+1} = \phi_{i,j}&(f(\beta_{1,j} \times h_{1,j},...,\beta_{i-1,j} \times h_{i-1,j}, h_{i,j});\\
&w_{i,j},b_{i,j}), i\in\{2,...,m\},j\in\{1,...,n\}
\end{split}
\end{equation}

\begin{equation}
\begin{split}
h_{i,out} = &\phi_{i,n+1}(f(\beta_{1,n+1} \times h_{1,n+1},...,\beta_{i,n+1} \times h_{i,n+1}\\
& ,h_{i,n+1});w_{i,n+1},b_{i,n+1}), i\in\{2,...,m\}
\end{split}
\end{equation}
here $h_{i,out}$ is the output of SGN-$i$,
and $\beta$ is a weight to control the proportion of different information,

\begin{equation}
\begin{split}
h_{1,1}^{'} = &\phi_{1,1}(f(\alpha_{1,out} \times h_{1,out},..., \alpha_{i,out} \times h_{i,out}, \\
&h_{i,in});w_{1,1},b_{1,1}),i\in\{1,...,m\}
\end{split}
\end{equation}
the above function is the first-step operation of back connection. Firstly, we alter the input of SGN-1 to receive all $h_{i,out}$'s and the original input $h_{1,in}$,

\begin{equation}
\begin{split}
h_{1,j+1}^{'} &= \phi_{1,j}(f(h_{1,j}^{'},\beta_{2,j} \times h_{2,j},...,\beta_{i,j} \times h_{i,j}); \\
&w_{1,j},b_{1,j}),i\in\{1,...,n\},j\in\{1,...,m\}
\end{split}
\end{equation}
and then it receives all single-granulation feature of each layer, $h_{1,j}^{'}$ represent the output of the $j-th$ layer in SGN-1 after back connection,

\begin{equation}
\begin{split}
h_{1,out}^{'} = &\phi_{1,n+1}(f(h_{1,n+1}^{'},...,\beta_{i,n+1} \times h_{i,n+1}); \\
&w_{1,n+1},b_{1,n+1}),i\in\{1,...,m\}
\end{split}
\end{equation}
where $h_{1,out}^{'}$ is the SGN-1 output after back connection,
Till now, we have articulated the workflow of our multi-granulation network.

\begin{figure}[htb]
\centering
 	\includegraphics[width=0.4\textwidth]{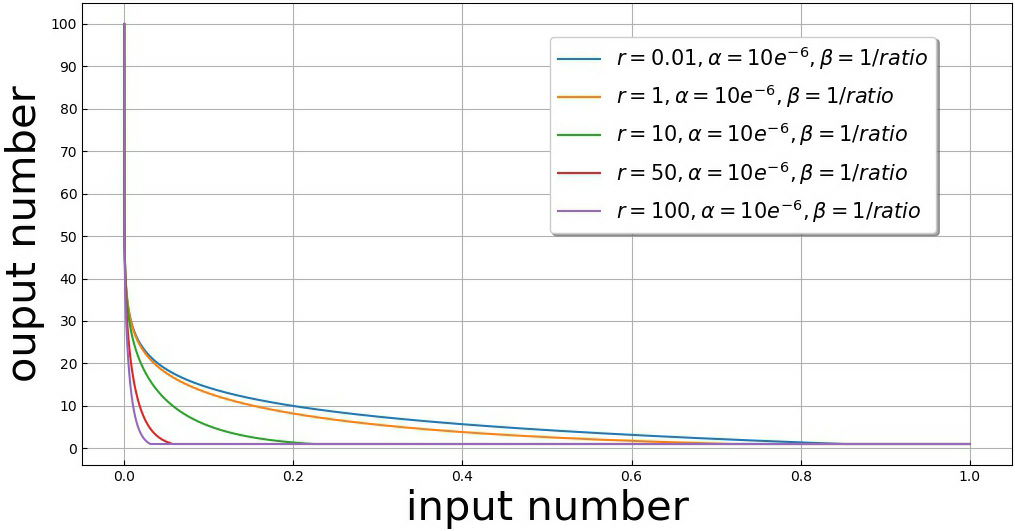}
 \includegraphics[width=0.4\textwidth]{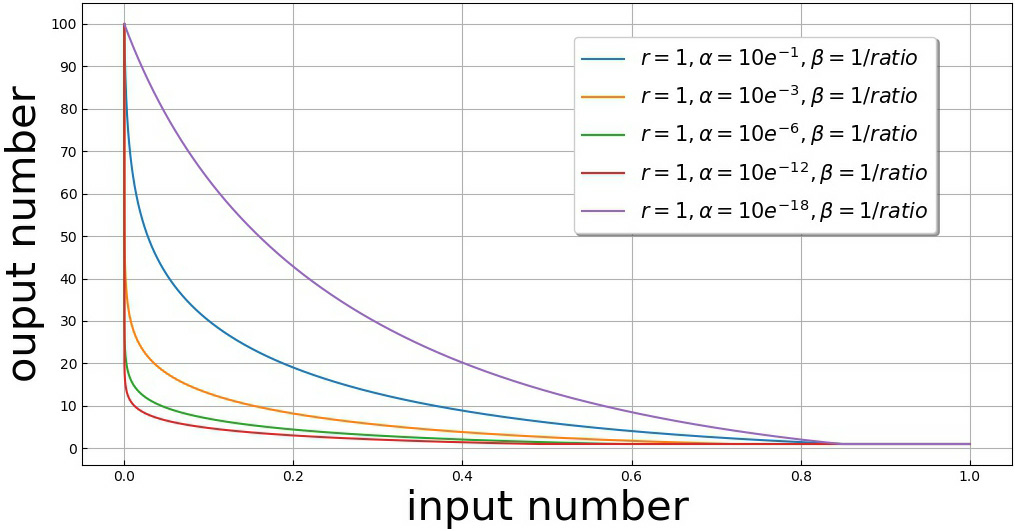}
 \includegraphics[width=0.4\textwidth]{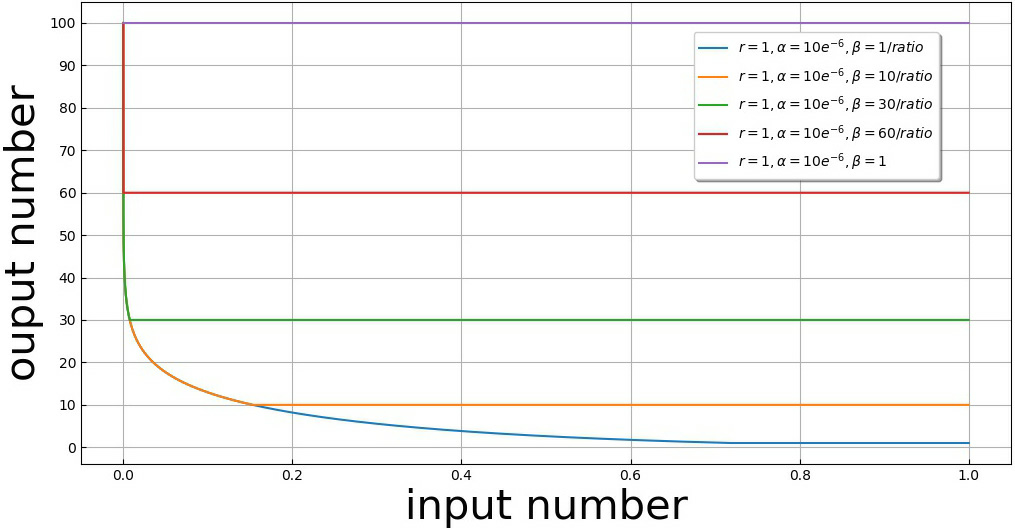}\\
    \caption{The effect of different parameters. }
    \label{fig:IMEF_parameters}
\end{figure}

\subsection{Instantiations}
\label{sec:Instantiations}

Next, we describe several versions of \emph{f($\cdot$)} and demonstrate through a series of experiments (\tablename~\ref{table:investigation framework}) that the bidirectional information flow is the main reason for the performance improvements.

\textbf{RMCN} Inspired by He \etal \cite{He2015Deep},
we consider \emph{f ($\cdot$)} in a form of residual connection, and set $\alpha $, $\beta$ to 1.
the function can be written as:
%
%

\begin{equation}
\begin{split}
f(x_{1},...,x_{n}) = \Sigma_{i}^{n}x_{i}
\end{split}
\end{equation}

\textbf{DMCN}  Following Huang \etal \cite{Huang2017Densely},
we also provide a version of concatenation with $\alpha = 1$. The weight $\beta$  is set as 1 in cooperation connection and $0$ in back connection.
This function is written as:

\begin{equation}
\begin{split}
f(x_{1},...x_{n}) = [x_{1},...x_{n}]
\end{split}
\end{equation}

The above two schemes are used in our multi-granulation cooperative neural network for information fusion.
An effective way of feature fusion is very important, and we believe that alternative versions can further improve the results.

\subsection{Raw Illumination Map Estimation Function}
\label{sec:IMEF}
Although numerous methods have been proposed to deal with the image enhancement \cite{Chen_2019_ICCV, dark} in extreme low-light environment, they hardly preserve desired high dynamic range.
The key reason is they choose to amplify raw data linearly  (\figurename~\ref{fig:IMEF}).
To make an improvement, we propose raw illumination map estimation function (RIMEF, as shown below in equation 1 ) to amplify raw data in a non-linear fashion:

\begin{equation}
\begin{split}
M &= max(m_{f},\beta)*ratio,\\
&~s.t.~~\beta \in [\emph{$\frac{1}{ratio}$}, 1]
\end{split}
\end{equation}
where \emph{$\beta$} is the lower bound of RIMEF, $ratio$ is the amplification ratio obtained by calculating the difference of the exposure between the input image and the reference image (e.g., x100, x250, or x300). It is observed that RIMEF is made up of two parts, an illumination map function \emph{$m_{f}$} that calculates illumination map and a maximum function \emph{max($\cdot$)} that guarantees M, the final result of the estimation function to be no less than 1, which avoids bad pixels in the enhanced image.

\begin{equation}
\begin{split}
&m_{f} = \frac{e^{-rx}*log(x+\alpha)}{log(\alpha)},\\
~s.t.~~&r \in (0,+\infty), \alpha \in (0,1), x \in [0,1]
\end{split}
\end{equation}

\begin{figure*}[htb]
\centering
\begin{tikzpicture}[zoomboxarray, zoomboxes below, zoomboxarray inner gap=0.1cm, zoomboxarray columns=2, zoomboxarray rows=2]
    \node [image node] { \includegraphics[trim=0 0 0 0, clip, width=1.1in]{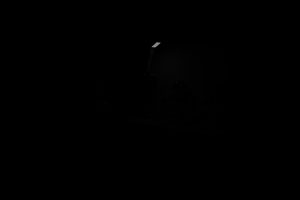} };
        \zoombox[magnification=1,color code=red]{0.58,0.52}
        \zoombox[magnification=4,color code=green]{0.3,0.53}
        \zoombox[magnification=4,color code=black]{0.5,0.70}
        \zoombox[magnification=4,color code=yellow]{0.58,0.50}
        \label{fig:visual_compare_IMEF_a}
\end{tikzpicture}
\begin{tikzpicture}[zoomboxarray, zoomboxes below, zoomboxarray inner gap=0.1cm, zoomboxarray columns=2, zoomboxarray rows=2]
    \node [image node] { \includegraphics[trim=0 0 0 0, clip, width=1.1in]{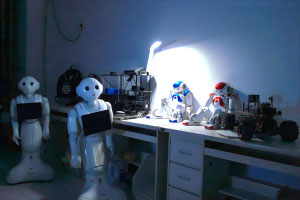} };
        \zoombox[magnification=1,color code=red]{0.58,0.52}
        \zoombox[magnification=4,color code=green]{0.3,0.53}
        \zoombox[magnification=4,color code=black]{0.5,0.70}
        \zoombox[magnification=4,color code=yellow]{0.58,0.50}
        \label{fig:visual_compare_IMEF_b}
\end{tikzpicture}
\begin{tikzpicture}[zoomboxarray, zoomboxes below, zoomboxarray inner gap=0.1cm, zoomboxarray columns=2, zoomboxarray rows=2]
    \node [image node] { \includegraphics[trim=0 0 0 0, clip, width=1.1in]{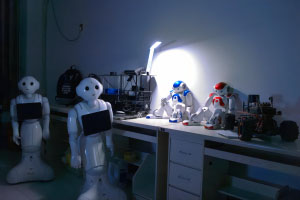} };
        \zoombox[magnification=1,color code=red]{0.58,0.52}
        \zoombox[magnification=4,color code=green]{0.3,0.53}
        \zoombox[magnification=4,color code=black]{0.5,0.70}
        \zoombox[magnification=4,color code=yellow]{0.58,0.50}
        \label{fig:visual_compare_IMEF_c}
\end{tikzpicture}
\begin{tikzpicture}[zoomboxarray, zoomboxes below, zoomboxarray inner gap=0.1cm, zoomboxarray columns=2, zoomboxarray rows=2]
    \node [image node] { \includegraphics[trim=0 0 0 0, clip, width=1.1in]{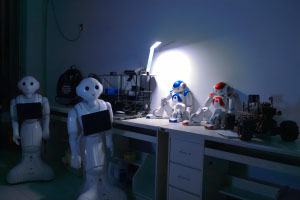} };
        \zoombox[magnification=1,color code=red]{0.58,0.52}
        \zoombox[magnification=4,color code=green]{0.3,0.53}
        \zoombox[magnification=4,color code=black]{0.5,0.70}
        \zoombox[magnification=4,color code=yellow]{0.58,0.50}
        \label{fig:visual_compare_IMEF_d}
\end{tikzpicture}
\begin{tikzpicture}[zoomboxarray, zoomboxes below, zoomboxarray inner gap=0.1cm, zoomboxarray columns=2, zoomboxarray rows=2]
    \node [image node] { \includegraphics[trim=0 0 0 0, clip, width=1.1in]{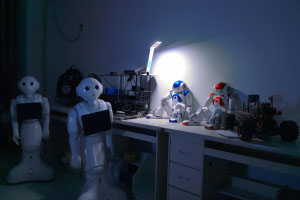} };
        \zoombox[magnification=1,color code=red]{0.58,0.52}
        \zoombox[magnification=4,color code=green]{0.3,0.53}
        \zoombox[magnification=4,color code=black]{0.5,0.70}
        \zoombox[magnification=4,color code=yellow]{0.58,0.50}
            \label{fig:visual_compare_IMEF_e}
\end{tikzpicture}
\begin{tikzpicture}[zoomboxarray, zoomboxes below, zoomboxarray inner gap=0.1cm, zoomboxarray columns=2, zoomboxarray rows=2]
    \node [image node] { \includegraphics[trim=0 0 0 0, clip, width=1.1in]{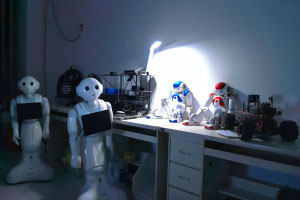} };
        \zoombox[magnification=1,color code=red]{0.58,0.52}
        \zoombox[magnification=4,color code=green]{0.3,0.53}
        \zoombox[magnification=4,color code=black]{0.5,0.70}
        \zoombox[magnification=4,color code=yellow]{0.58,0.50}
    \label{fig:visual_compare_IMEF_f}
\end{tikzpicture}\\
\begin{tabular}{cx{4cm}cx{4cm}cx{4cm}cx{4cm}cx{4cm}cx{4cm}}
~~~~~~~~~~~~(a) & ~~~~~~~~(b) & ~~~~~~~(c) &~~~~~~~(d) & ~~~~~~~(e) & ~~~~~(f)
\end{tabular}
\caption{Visual comparison of RIMEF with different parameters.
(a) Raw data.
(b) LTS \cite {dark} with \emph{$ratio=200$}.
(c) DMCN-3 with \emph{$r=10,\alpha=10^{-6},\beta=\frac{1}{ratio},ratio=300$}.
(d) DMCN-3 with \emph{$r=1,\alpha=10^{-6},\beta=\frac{1}{ratio},ratio=200$}.
(e) DMCN-3 with \emph{$r=5,\alpha=10^{-6}, \beta=\frac{1}{ratio}, ratio=200$}.
(f) DMCN-3 with \emph{$r=1,\alpha=10^{-1}, \beta=\frac{1}{ratio}, ratio=200$}.}
\label{fig:visual_compare_IMEF}
\end{figure*}

In particular, RIMEF can accommodate to different levels of exposure.
Like \emph{$\beta$}, the parameters \emph{$r$} and \emph{$\alpha$} are set to control the RIMEF, which affects the exposure of the image.
The \emph{$\beta$} is set as $1$ during network training, and as $\frac{1}{ratio}$ when  processing extreme low-light HDR raw image.
We empirically set the \emph{$\alpha=10^{-6}$} and \emph{$r=1$}.
The visual effects are presented in \figurename~\ref{fig:visual_compare_IMEF}.  \figurename~\ref{fig:visual_compare_IMEF}(b)
 shows that when RIMEF degenerates to linear magnification (M = ratio), the image will suffer from a serious over-exposure.

\label{sec:loss function}
\begin{figure}[htp]
  	\begin{tabular}{cccc}
  		\centering
     	\includegraphics[width=1.55in]{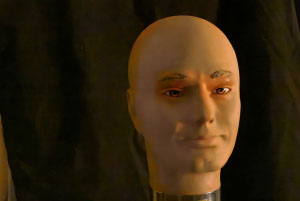}&
     	\includegraphics[width=1.55in]{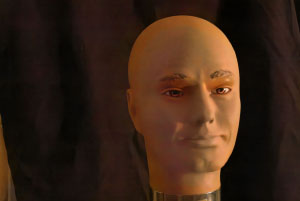} \\
     (a) LTS \cite{dark}& (b) RMCN w/o \emph{$\ell_{s}$} \\
     	\includegraphics[width=1.55in]{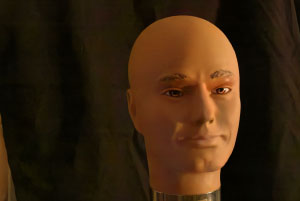} &
     	\includegraphics[width=1.55in]{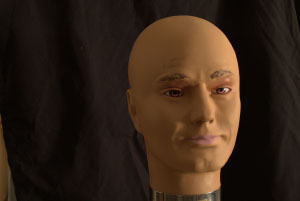} \\
        (c) RMCN w/ \emph{$\ell_{s}$} &(d) Ground truth \\
  	\end{tabular}
    	\caption{Visual effect with different loss.}
  \label{fig:visual_loss}
\end{figure}

\subsection{Loss Function}

We design a novel multi-granulation loss {$\ell_{mu}$} and minimize it during training:
\begin{equation}
\begin{split}
\ell_{mu} = \frac{1}{K}\sum_{\kappa=1}^{K}\lambda_{r}\ell_{r}^{\kappa} + \lambda_{s}\ell_{s}^{\kappa}
\end{split}
\end{equation}

where \emph{$K$} is the number of image pairs, \emph{$\lambda_{r}$} and \emph{$\lambda_{s}$} are the corresponding weights.
We empirically set \emph{$\lambda_{r}$}=1 and \emph{$\lambda_{s}^{i}$}=1,
where \emph{$\ell_{r}$} and \emph{$\ell_{s}$} are the image reconstruction loss and image smoothness loss respectively.

\textbf{Image Reconstruction Loss}
The image reconstruction loss is written as:

\begin{equation}
\ell_{r} = \sum_{i=2}^{N}|h_{i,out}-I|+|h_{1,out}^{'}-I|
\end{equation}

Here \emph{$N$} is the number of SGNs in the MCN,
\emph{$h_{i,out}$} is the \emph{$i-th$} network's output, \emph{$I$} is ground-truth image, \emph {$h_{1,out}^{'}$} is the result of SGN-1 after back connection.

\textbf{Image Smoothness Loss}
In a natural image the illumination is usually locally smooth as it has a large spatial difference in low-light image.
Inspired by Xu \etal \cite{xu2012structure},
we design a smoothness loss based-on  total variation.
The visual effects are shown in \figurename~\ref{fig:visual_loss} and the loss function is defined as:

\begin{equation}
\ell_{s} = \sum_{i=2}^{N}|\partial h_{i,out}|+|\partial h_{1,out}^{'}|
\end{equation}

where the {$\partial$} is the total variation in the x and y directions.

\section{Dark High Dynamic Range Dataset}

\begin{figure}[htp]
  	\begin{tabular}{cccc}
  		\centering
     	\includegraphics[width=1.55in]{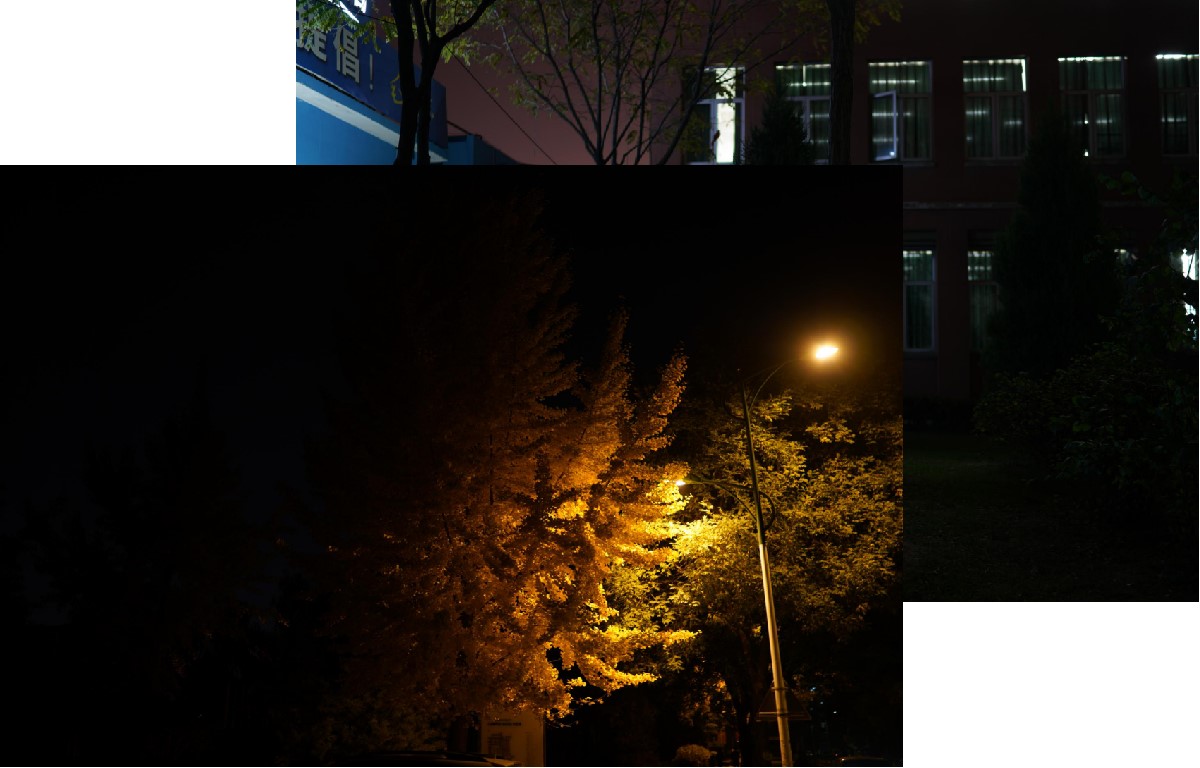}&
     	\includegraphics[width=1.55in]{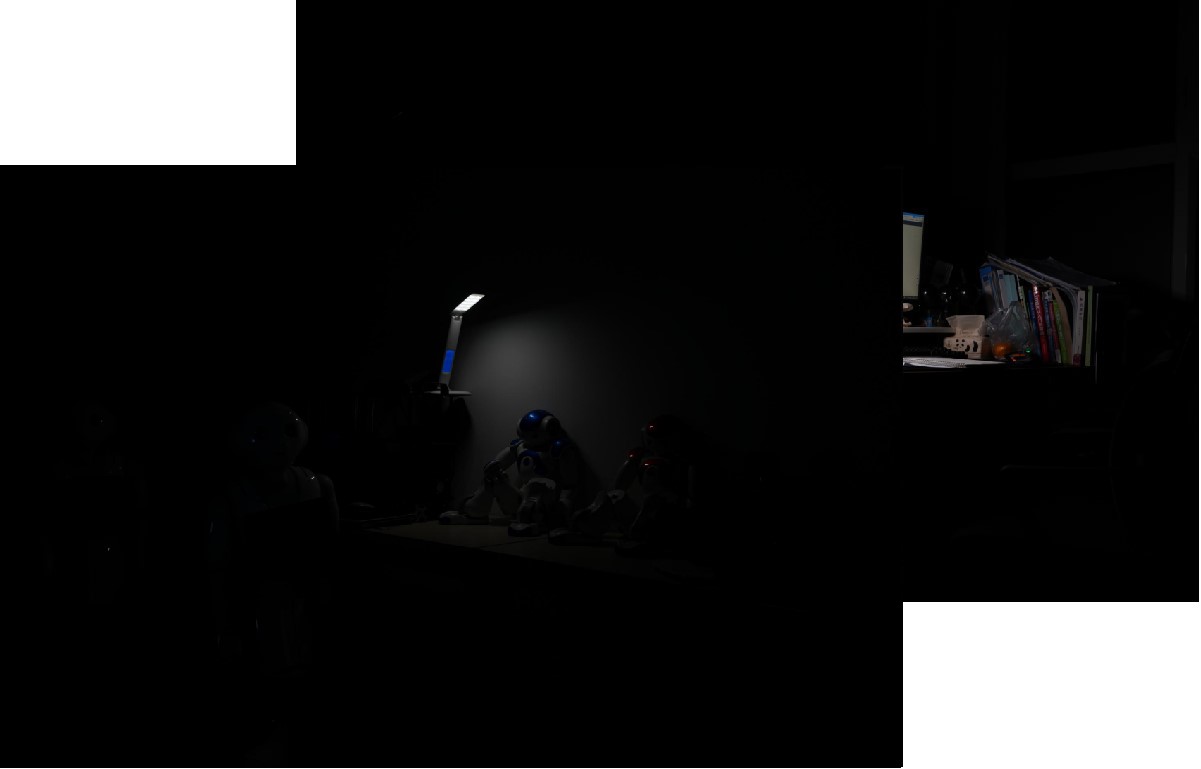} \\
        (a) outdoor & (b) indoor  \\
  	\end{tabular}
    	\caption{Some samples of our DHDR dataset .}
  \label{fig:dataset}
\end{figure}

To the best of our knowledge, there exist few raw datasets for the exclusive study on the high dynamic preservation of raw image in low-light environment. To advance the research progress, we collect a new dataset using camera Sony A7R3,  called \textbf{Dark High Dynamic Range (DHDR)} dataset for evaluating the performance of high dynamic preservation algorithm.
The dataset includes 30 raw images in low-light high dynamic environment where exposure time is set as $\frac{1}{25}$, and 10 images among them are from indoor scenes and the 20 others from outdoor scenes.
Considering the difficulty of obtaining ground truth, our dataset can be used for visual evaluation barely.
Here are some samples of reference images shown in \figurename~\ref{fig:dataset}.

\section{Experimental Analysis}

First of all, we introduce the dataset and training details in Section~\ref{sec:Training details and parameters}.
Secondly, we conduct visual analysis and quantitative analysis in Section~\ref{sec:Visual and Quantitative Comparisons}.
In Section~\ref{sec:Investigation of Illumination Map Estimation Function}, we further analyze the raw illumination map estimation function in depth.
In Section~\ref{sec:Investigation of coorNet}, we conduct more comprehensive analysis of our multi-granulation cooperative network.
Finally, we will compare the performance between different network structures in Section~\ref{sec:Investigation of Framework}.

\subsection{Dataset and Training Details}
\label{sec:Training details and parameters}
A series of experiments are performed on See-in-the-Dark (SID) \cite{dark} dataset, which includes images shot by Sony $\alpha$ 7sII (Sony subset) and Fujifilm X-T2 (Fuji subset).
SID dataset contains indoor and outdoor scenes.
We use Sony dataset and Fuji dataset for training the network, respectively. All experiments has been carried out on single Tesla P100 GPU.  Following experiment setup of \cite{dark}, we trained our proposed multi-granularity cooperative networks.
The input of our network is short-exposed raw data while the ground truth is the corresponding long-exposure image. To enhance the performance of our network, we randomly crop a 512 $\times$ 512 patch with random flipping and rotation.
We choose learning rate to be 10$^{-4}$ at the beginning and 10$^{-5}$ after 2000 epochs.
For optimization we use Adam \cite{Kingma2014Adam} and default parameters.
Epoch is set as 4000 times.

\begin{table}[htp]
\begin{center}
\setlength{\tabcolsep}{6pt}
\begin{tabular}{l c c  p{3mm} c c }
\toprule
&\multicolumn{2}{c}{Sony subset} &&  \multicolumn{2}{c}{Fuji subset} \\
\midrule

\rowcolor{color3}& PSNR & SSIM && PSNR & SSIM     \\
\midrule
\multicolumn{1}{l}{LTS} &
28.639  & 0.768 & &
 26.477 & 0.688  \\
\multicolumn{1}{l}{RMCN-3} &
29.146& 0.775 &&
\textbf{26.885} & 0.696 \\
\multicolumn{1}{l}{DMCN-3} &
\textbf{29.215} & \textbf{0.778}  & &
26.837 & \textbf{0.699}\\
\bottomrule
\end{tabular}
\end{center}\vspace{-0.5em}
\caption{
Quantitative comparison with the state-of-the-art method LTS \cite{dark} on SID dataset.
Best results is in bold. All results in the table is the average performance with several  random seed.}
\label{table:compare_quantitative}
\end{table}

\begin{figure}[htb]
\begin{tikzpicture}[zoomboxarray, zoomboxes below, zoomboxarray inner gap=0.1cm, zoomboxarray columns=2, zoomboxarray rows=2]
    \node [image node] { \includegraphics[trim=0 0 0 0, clip, width=1in]{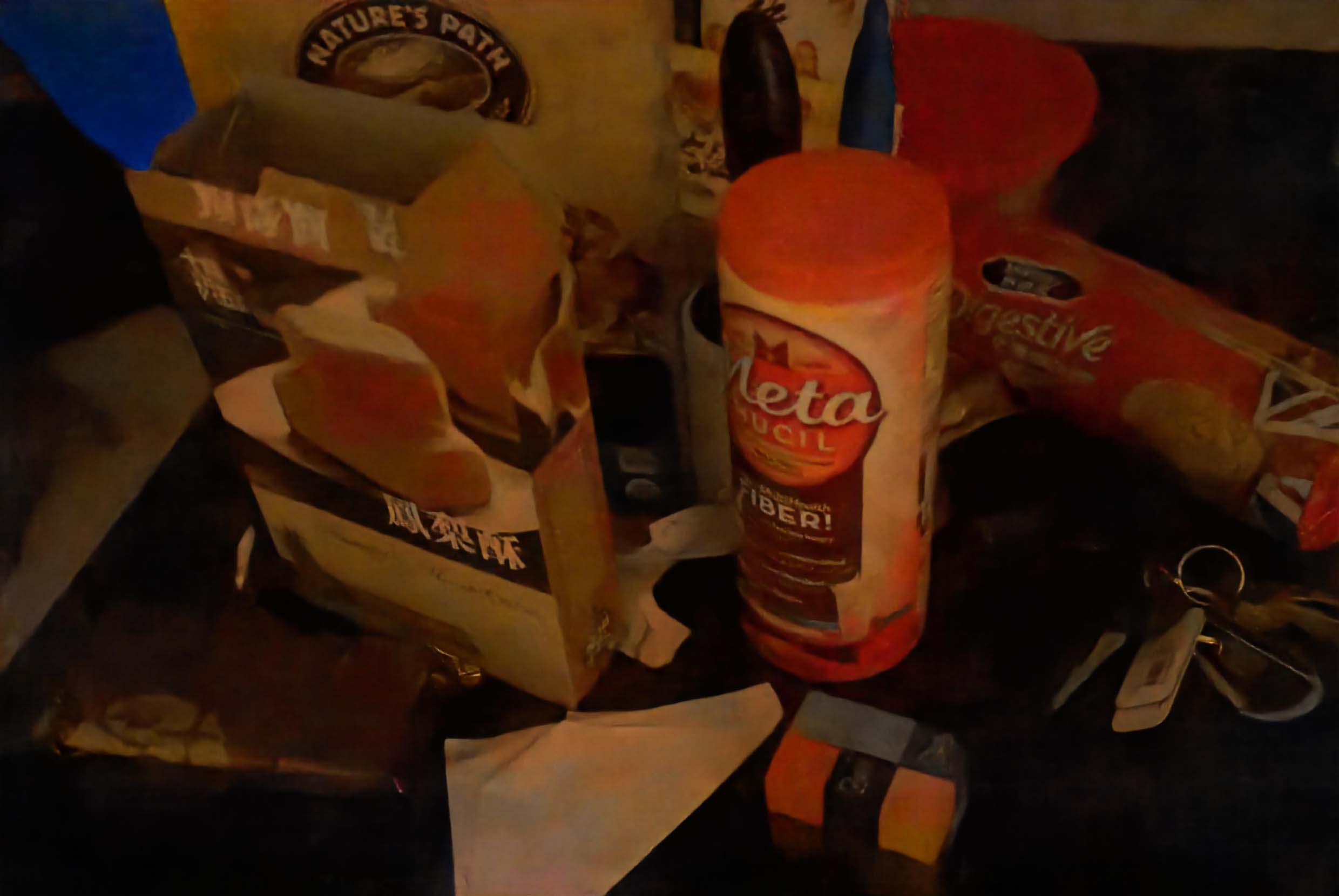} };
    \zoombox[magnification=6,color code=red]{0.41,0.25}
    \zoombox[magnification=6,color code=green]{0.35,0.41}
    \zoombox[magnification=6,color code=yellow]{0.91,0.3}
    \zoombox[magnification=6,color code=blue]{0.21,0.55}
\end{tikzpicture}
\begin{tikzpicture}[zoomboxarray, zoomboxes below, zoomboxarray inner gap=0.1cm, zoomboxarray columns=2, zoomboxarray rows=2]
    \node [image node] { \includegraphics[trim=0 0 0 0, clip, width=1in]{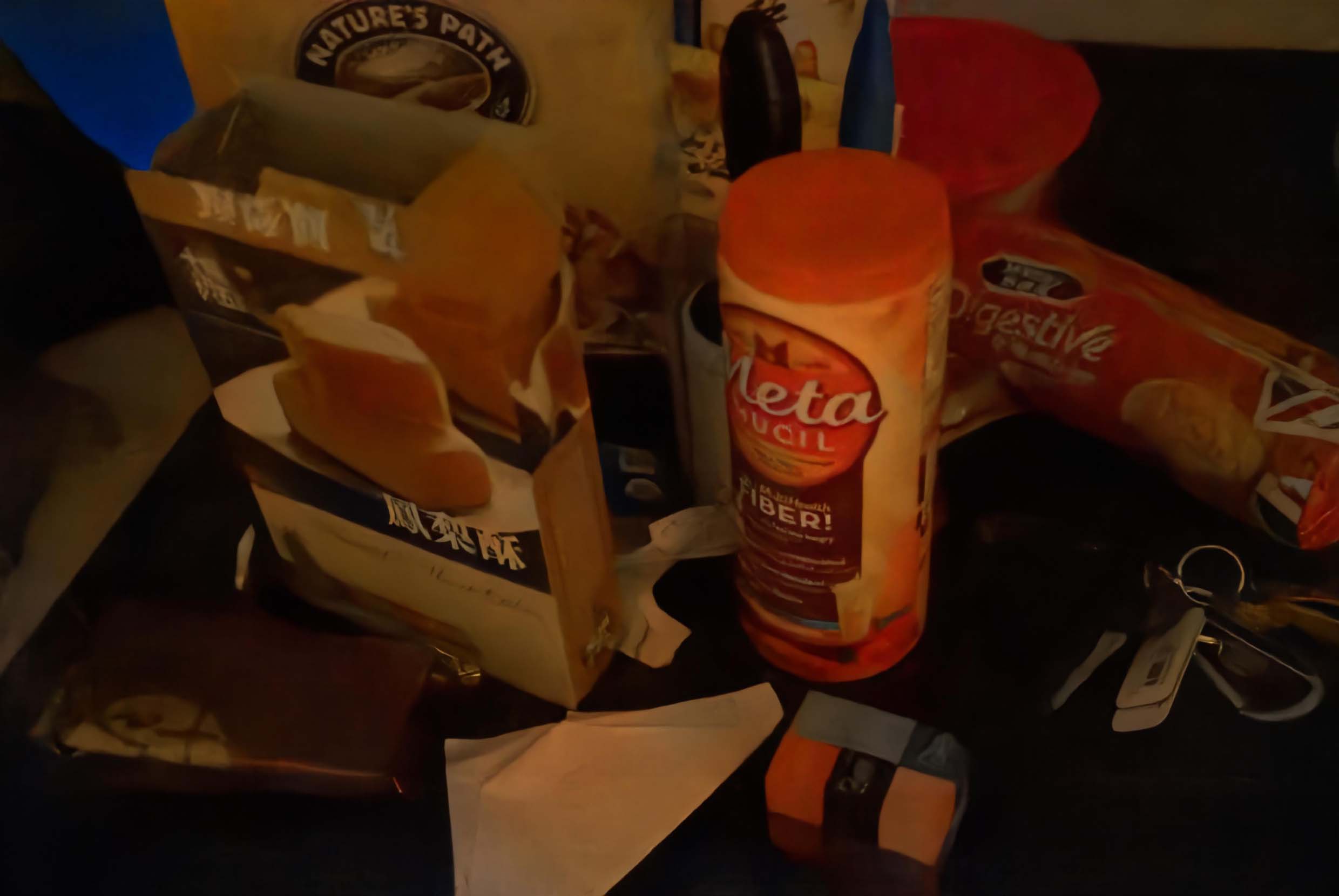} };
    \zoombox[magnification=6,color code=red]{0.41,0.25}
    \zoombox[magnification=6,color code=green]{0.35,0.41}
    \zoombox[magnification=6,color code=yellow]{0.91,0.3}
    \zoombox[magnification=6,color code=blue]{0.21,0.55}
\end{tikzpicture}
\begin{tikzpicture}[zoomboxarray, zoomboxes below, zoomboxarray inner gap=0.1cm, zoomboxarray columns=2, zoomboxarray rows=2]
    \node [image node] { \includegraphics[trim=0 0 0 0, clip, width=1in]{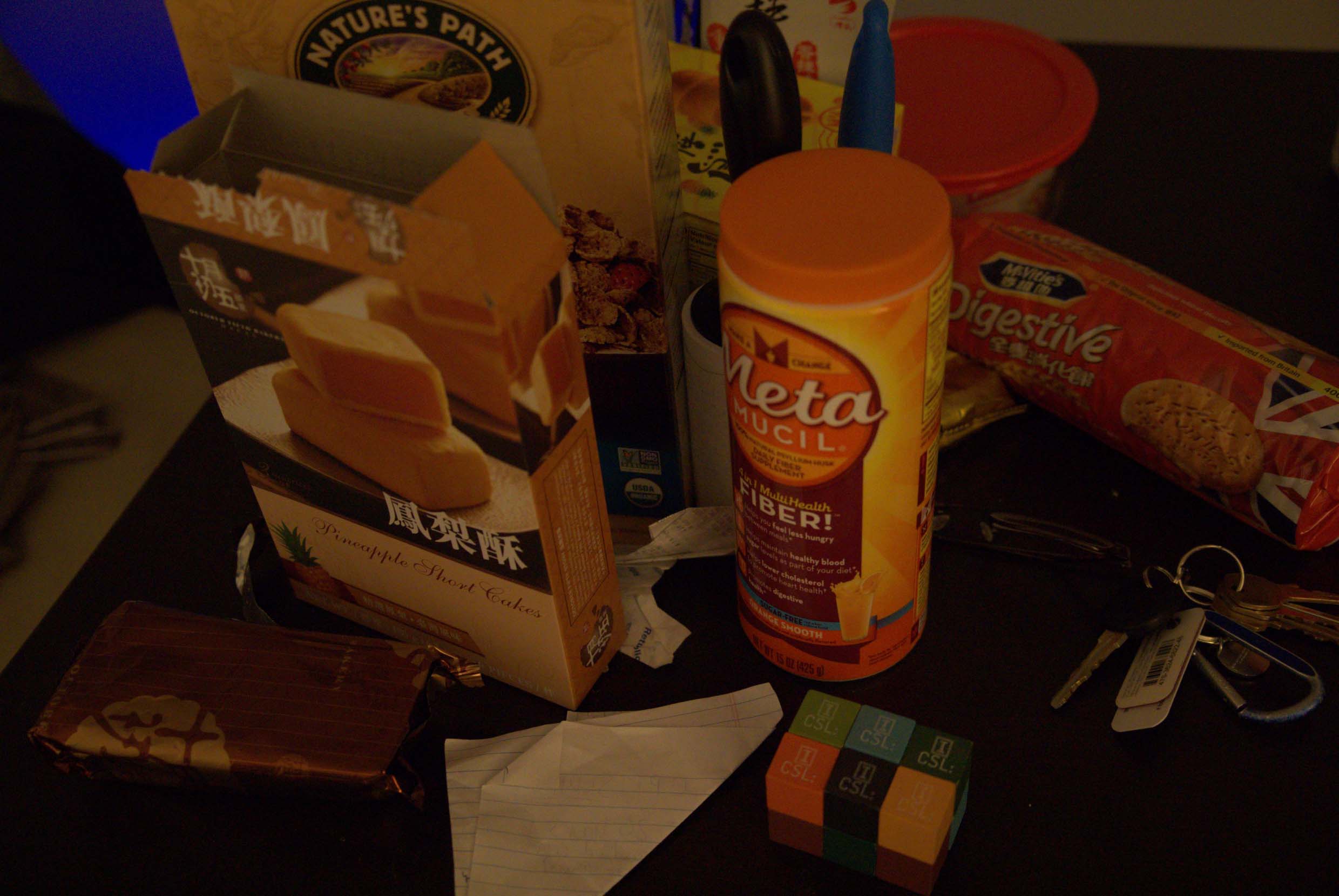} };
    \zoombox[magnification=6,color code=red]{0.41,0.25}
    \zoombox[magnification=6,color code=green]{0.35,0.41}
    \zoombox[magnification=6,color code=yellow]{0.91,0.3}
    \zoombox[magnification=6,color code=blue]{0.21,0.55}
\end{tikzpicture}
\begin{tabular}{ cx{3cm}cx{3cm}cx{3cm}}
	~~~~~~~~~~(a) &~~~~~~~~~~~~~~~~~~(b) & ~~~~~~~~~~~~~~~~~(c)  \\
\end{tabular}
\begin{tikzpicture}[zoomboxarray, zoomboxes below, zoomboxarray inner gap=0.1cm, zoomboxarray columns=2, zoomboxarray rows=2]
    \node [image node] { \includegraphics[trim=0 0 0 0, clip, width=1in]{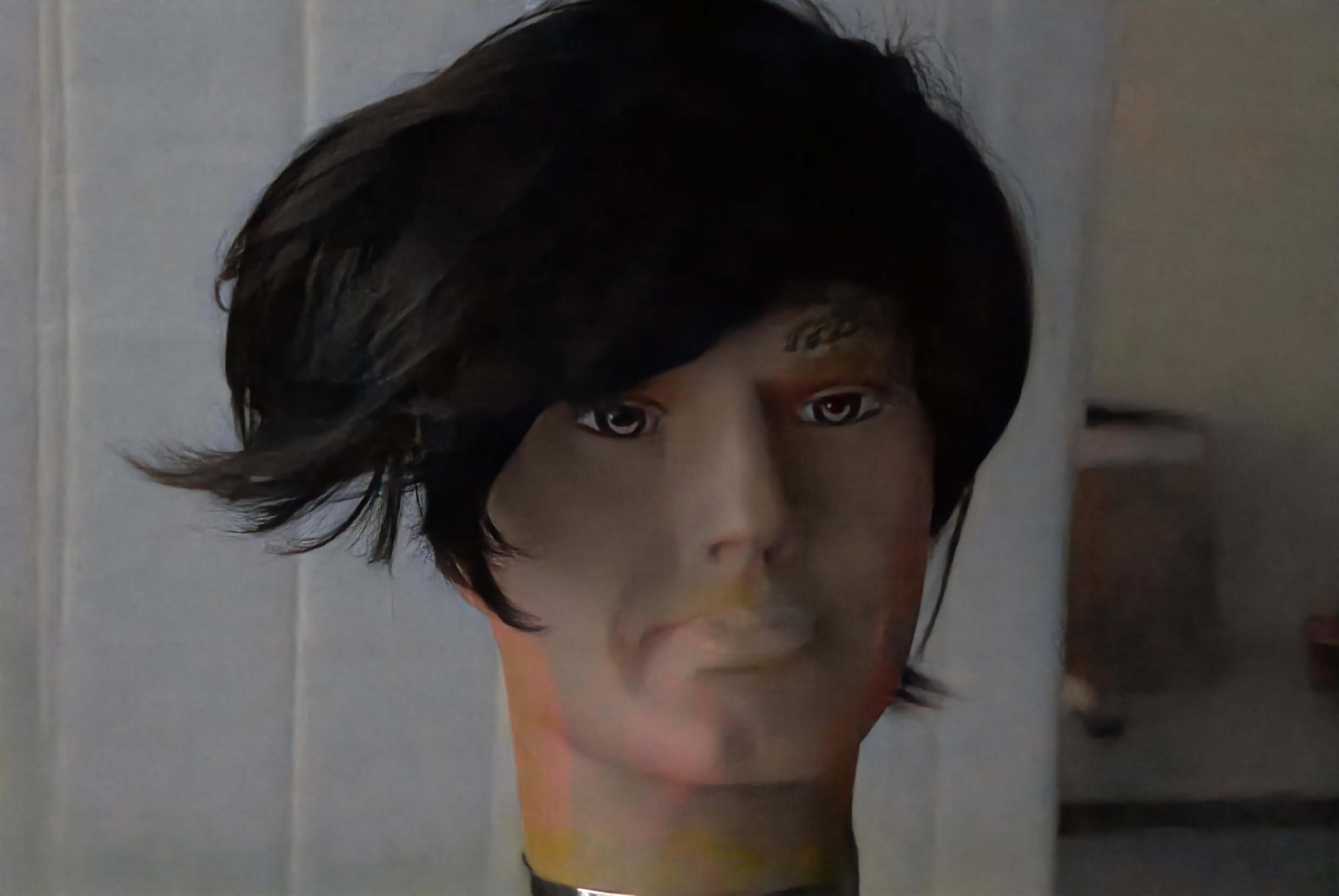} };
    \zoombox[magnification=4,color code=red]{0.55,0.33}
    \zoombox[magnification=4,color code=green]{0.45,0.55}
    \zoombox[magnification=4,color code=black]{0.43,0.08}
    \zoombox[magnification=4,color code=blue]{0.6,0.1}
\end{tikzpicture}
\begin{tikzpicture}[zoomboxarray, zoomboxes below, zoomboxarray inner gap=0.1cm, zoomboxarray columns=2, zoomboxarray rows=2]
    \node [image node] { \includegraphics[trim=0 0 0 0, clip, width=1in]{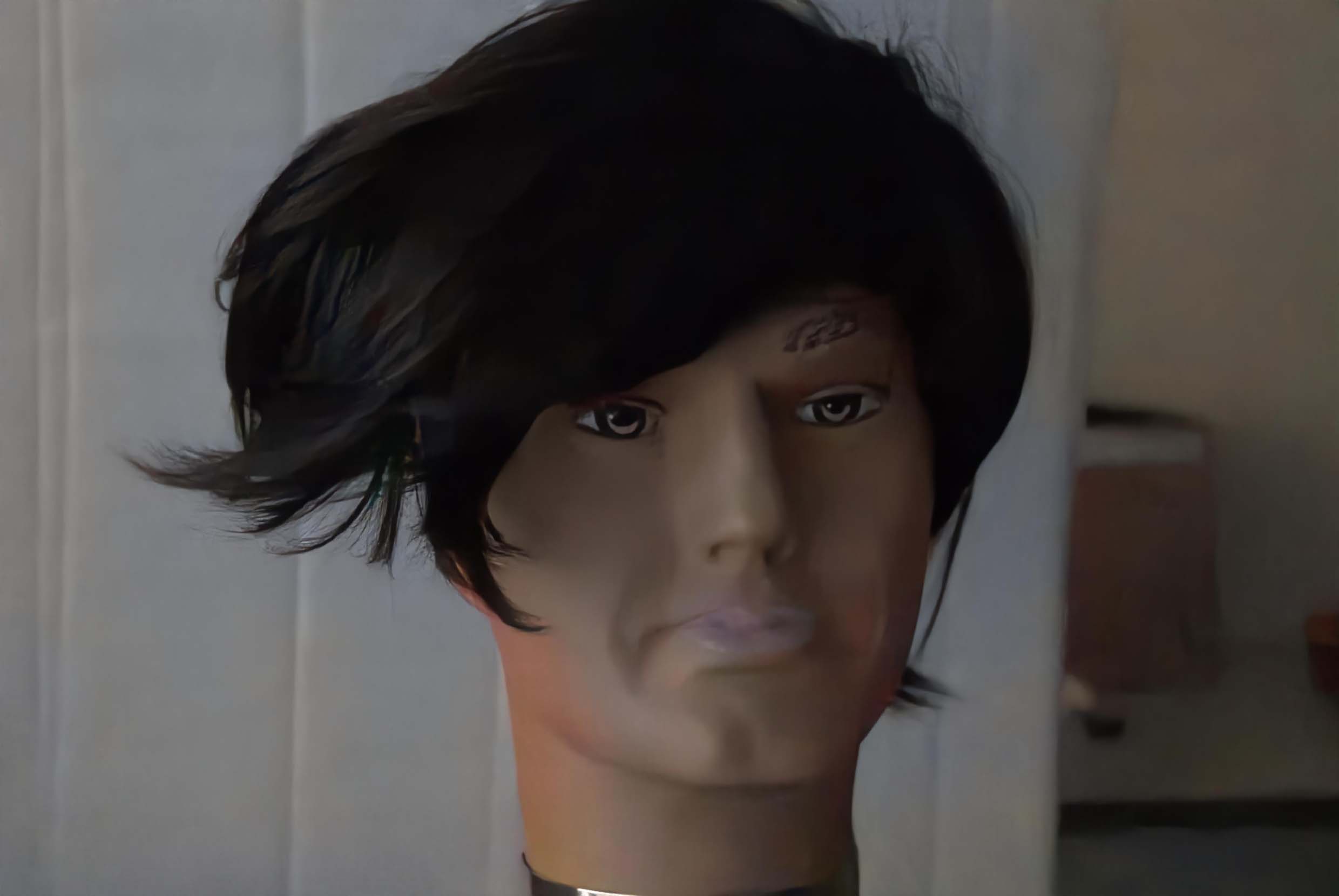} };
    \zoombox[magnification=4,color code=red]{0.55,0.33}
    \zoombox[magnification=4,color code=green]{0.45,0.55}
    \zoombox[magnification=4,color code=black]{0.43,0.08}
    \zoombox[magnification=4,color code=blue]{0.6,0.1}
\end{tikzpicture}
\begin{tikzpicture}[zoomboxarray, zoomboxes below, zoomboxarray inner gap=0.1cm, zoomboxarray columns=2, zoomboxarray rows=2]
    \node [image node] { \includegraphics[trim=0 0 0 0, clip, width=1in]{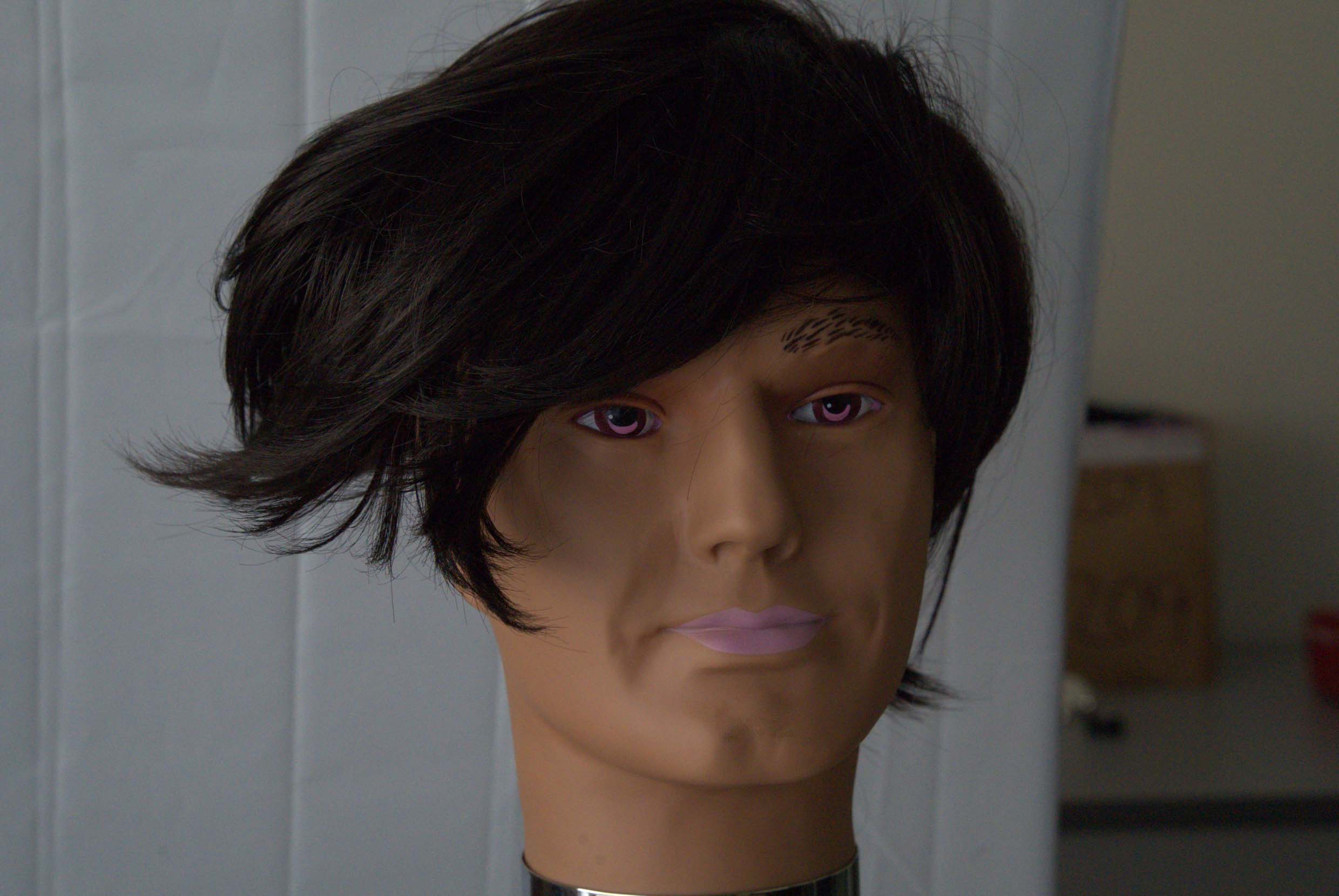} };
    \zoombox[magnification=4,color code=red]{0.55,0.33}
    \zoombox[magnification=4,color code=green]{0.45,0.55}
    \zoombox[magnification=4,color code=black]{0.43,0.08}
    \zoombox[magnification=4,color code=blue]{0.6,0.1}

\end{tikzpicture}
\begin{tabular}{ cx{4cm}cx{4cm}cx{4cm}}
	~~~~~~~~~~(d) &~~~~~~~~(e) & ~~~~~~(f)  \\
\end{tabular}
\caption{Visual comparison of extreme low-light image enhancement. Reference Ground Truth image (c,f) with corresponding method of LTS (a, d) and RMCN (b, e).}

\label{fig:final_visual_a}
\end{figure}
\begin{figure*}[htp]
\centering
  	\begin{tabular}{ccccccc}

     	\includegraphics[width=1.5in]{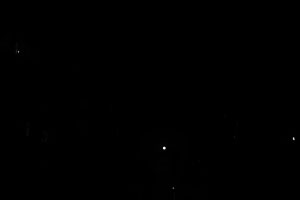}&
        \includegraphics[width=1.5in]{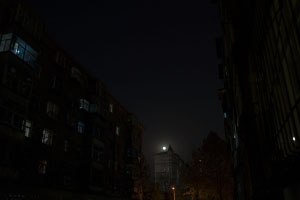}&
     	\includegraphics[width=1.5in]{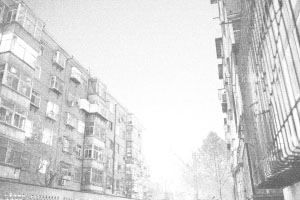} &
     	\includegraphics[width=1.5in]{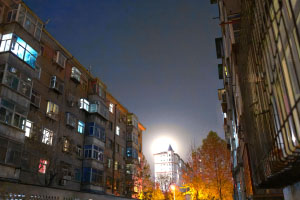} \\
      	(a)& (b)  & (c)& (d)    \\
  	\end{tabular}
    \begin{tabular}{cccc}
     	\includegraphics[width=1.5in]{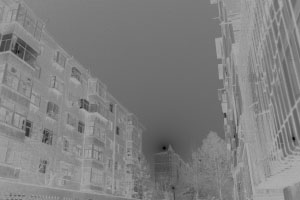}&
     	\includegraphics[width=1.5in]{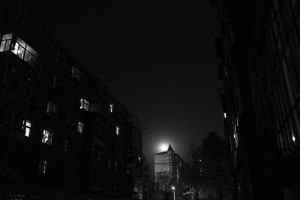} &
        \includegraphics[width=1.5in]{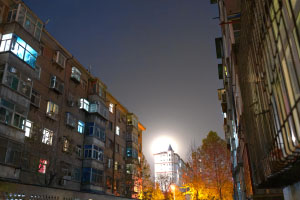}&
        \includegraphics[width=1.5in]{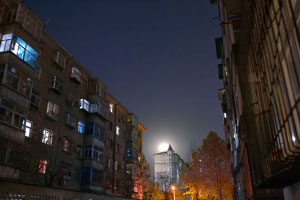}\\
       	(e)& (f) & (g) & (h)  \\
  	\end{tabular}
\caption{
  (a) Raw data.
  (b) Photoshop Camera Raw automatic mode processing.
  (c) Raw data amplified linearly.
  (d) LTS \cite{dark}.
  (e) Raw illumination map.
  (f) Raw data amplified by raw illumination map .
  (g) RMCN-3 w/o RIMEF.
  (h) RMCN-3 w/ RIMEF.}
\label{fig:IMEF}
\end{figure*}

\begin{table*}[htb]
\centering
\label{tab:perceptual}
\begin{adjustbox}{max width=1\textwidth}
\centering
\begin{tabular}{|l|l|l|l|l|l|l|l|}
\hline
Sony subset & ~~~RMCN-2& ~~~RMCN-3&~~~DMCN-3& RMCN-3 w/o BC& DMCN-3 w/o BC& ~~~RMCN-4& ~~~RMCN-5 \\
\hline
~~~~SGN-1 & ~~6.608 / 0.188& ~1.072 / 0.146& ~9.420 / 0.160& ~~~28.663 / 0.770& ~~~28.667 / 0.770&~-8.835 / 0.097& -15.024 / 0.080 \\
\hline
~~~~SGN-2 & \textbf{28.955 / 0.774}& 28.943 / 0.773& 28.944 / 0.772& ~~~28.851 / 0.771& ~~~28.905 / 0.773&  ~28.757 / 0.772& ~28.196 / 0.747\\
\hline
~~~~SGN-3 & ~~~~~~~~~~-& 29.086 / 0.775& 29.193 / 0.776& ~~~28.957 / 0.772& ~~~29.059 / 0.775& ~28.962 /0.775& ~28.493 / 0.749  \\
\hline
~~~~SGN-4 & ~~~~~~~~~~-& ~~~~~~~~~~-& ~~~~~~~~~~-& ~~~~~~~~~~~~~-& ~~~~~~~~~~~~-& ~29.084 / 0.775& ~28.622 / 0.750  \\
\hline
~~~~SGN-5 & ~~~~~~~~~~-& ~~~~~~~~~~-& ~~~~~~~~~~-& ~~~~~~~~~~~~~-& ~~~~~~~~~~~~-& ~~~~~~~~~~~~~-& ~\textbf{28.670 / 0.750} \\
\hline
~~~~SGN-1$'$& 28.898 / 0.774& \textbf{29.146 / 0.775}& \textbf{29.215 / 0.778}& ~~~~~~~~~~~~~-& ~~~~~~~~~~~~-& ~\textbf{29.105 / 0.774}& ~28.156 / 0.0.737 \\
\hline
~~~~Params& ~~~~~~~16M& ~~~~~~~23M& ~~~~~~~42M& ~~~~~~~~~~23M& ~~~~~~~~~42M& ~~~~~~~~~31M& ~~~~~~~~~39M \\
\hline
\end{tabular}
\end{adjustbox}
\caption{
Analysis of multi-granulation network.
Highest calculated measures (PSNR/ SSIM) in bold.
RMCN-3 w/o BC and DMCN-3 w/o BC is the result without back connection (BC).
}
\label{table:dsgr}
\end{table*}
\subsection{Visual and Quantitative Comparisons}
\label{sec:Visual and Quantitative Comparisons}
In this part, we compare our method on SID dataset with the state-of-the-art method Chen \etal \cite{dark} and name it as LTS in the following.
RMCN-3 is a residual version of multi-granulation cooperative network and DMCN-3 is a dense version of multi-granulation cooperative network, which both contain three single-granulation networks.
\tablename~\ref{table:compare_quantitative} shows quantitative results and we can see our method DMCN-3 gets the best result and RMCN-3 gets the second result.
At the same time, we compare their visual effects and pose them in \figurename~\ref{fig:final_visual_a}.
The figure indicates that our result has less noise, smoother lines, and more realistic colors,
demonstrating the great generalization ability of our network structure by the comparison of quantitative and visual results.
Note that as the test environment changes, the result of the trained model by Chen \etal \cite{dark} we use can not identically amount to the one from the original paper.

\subsection{Investigation Raw Illumination Map Estimation Function}
\label{sec:Investigation of Illumination Map Estimation Function}
In this section, the property of RIMEF is studied on the DHDR dataset.
As can be seen from \figurename~\ref{fig:final_visual_a} (b), the moon and street lights are well-exposed, but the rest of the picture is badly under-exposed.
\figurename~\ref{fig:final_visual_a} (c) and \figurename~\ref{fig:final_visual_a}  (d) give the processing results of linearly scaling the raw data.
It can be seen that the under-exposed part of the whole image is largely improved.
However, the moon and street lamp are over-exposed.
In \figurename~\ref{fig:final_visual_a}  (e), the raw data is scaled in a non-linear way using RIMEF. Therefore, the image enhanced by our method is most similar to human perception, despite the slight over-exposure phenomenon around the moon.

\subsection{Investigation Multi-granulation Cooperative Network}
\label{sec:Investigation of coorNet}
\begin{figure}[htp]
\centering
  	\begin{tabular}{cccc}
  		\centering
     	\includegraphics[width=1.55in]{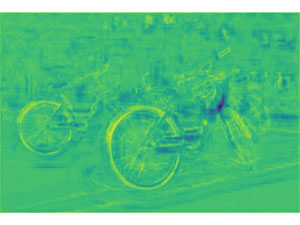}&
     	\includegraphics[width=1.55in]{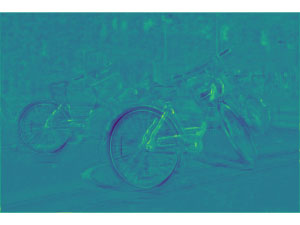} \\
        (a) SGN-1 &(b) SGN-2\\
     	\includegraphics[width=1.55in]{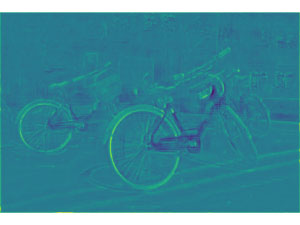} &
     	\includegraphics[width=1.55in]{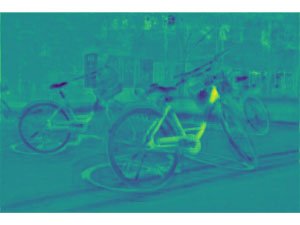} \\
        (c) SGN-3 & (d) SGN-1$'$  \\
  	\end{tabular}
  	\caption{Visual effect of convolution feature.}
  \label{fig:feature}
\end{figure}

In this part, we mainly study MCN.
From the fourth and fifth columns of Table \ref{table:dsgr}, we see that bidirectional information flow is a key factor to improve network performance.
From the second and third columns, we see that the way of information fusion between different SGNs will have a certain impact on the performance of MCN.
By observing other columns, it can be found that the number of SGN has a great influence on the performance of MCN, among which three SGN networks are the best choice.
When the number of SGN increases to four or five, the performance of MCN will degrade.
This is a interesting phenomenon.
In the future, this work will continue to study the causes of this phenomenon and how to cooperate more efficiently between networks.
The visual results of the convolution feature in RMCN-3 are shown in \figurename~\ref{fig:feature}, which can demonstrate the feature information will more abundant through multiple SGN cooperation.

\subsection{Investigation Structure}
\label{sec:Investigation of Framework}
To evaluate the learning effectiveness and generalization ability of our neural network framework in terms of depth, we quantitatively compare our method with two network structures ResNet and DenseNet.
First, we simply triple the number of convolutional layers in each convolution block and leave the rest unchanged.
Then the residual connection and dense connection are added to the expanded network respectively.
The above three nets are named Bignet, ResBignet and DenseBignet.
We use Sony sub-dataset to train these nets under the same training strategy and test them in the same environment, the results are reported in \tablename~\ref{table:investigation framework}.

\begin{table}[htb]
\centering

\label{tab:perceptual}
\begin{adjustbox}{max width=0.66\textwidth}
\centering
\begin{tabular}{|l|l|l|l|l|}
\hline
Sony subset & PSNR& SSIM & Params\\
\hline
LTS  & 28.639& 0.768 & 8M \\
\hline
BigNet  & 28.067& 0.760 & 23M\\
\hline
ResBigNet  & 28.682& 0.768 & 23M \\
\hline
DenseBigNet & 27.811& 0.748 & 78M \\
\hline
RMCN-3  & 29.146& 0.775 &23M  \\
\hline
DMCN-3 & \textbf{29.215}& \textbf{0.778} &41M\\
\hline
\end{tabular}
\end{adjustbox}
\caption{
Analysis of different network structure.
Highest calculated measures in bold.
}
\label{table:investigation framework}
\end{table}


\section{Conclusion}
\label{sec:Conclusion}
In this paper, a novel network architecture has been proposed for processing extreme low-light images and correspondingly the loss function has been modified for our task and
the raw illumination map estimation function (RIMEF) is designed to preserve high dynamic range in low-light environment.
Furthermore, we have extensively analyzed the function of the proposed network architecture and RIMEF before claiming the superiority of our method over the state-of-the-art one.

\bibliographystyle{plain}%

\bibliography{egpaper_for_review}

\end{document}